\documentclass[journal=jacsat,manuscript=article]{achemso}

\usepackage[version=3]{mhchem} 
\usepackage{graphicx}
\usepackage{dcolumn}
\usepackage{bm}
\usepackage{amssymb}
\usepackage[utf8]{inputenc}
\usepackage[T1]{fontenc}
\usepackage{mathptmx}
\usepackage{etoolbox}
\usepackage{color} 
\usepackage{lettrine}
\usepackage{url}
\usepackage[colorlinks = true,
            linkcolor = blue,
            urlcolor  = blue,
            citecolor = blue,
            anchorcolor = blue]{hyperref}
\usepackage[dvipsnames]{xcolor}
\usepackage{lineno}

\author{Obed \textbf{Alves Santos}}
\email{oa330@cam.ac.uk}
\affiliation[University of Groningen]
{Physics of Nanodevices, Zernike Institute for Advanced Materials, University of Groningen, Nijenborgh 4, Groningen, AG 9747, The Netherlands}
\alsoaffiliation[University of Cambridge]
{Cavendish Laboratory, University of Cambridge, Cambridge, CB3 0HE, United Kingdom}
\author{Bart J. \textbf{van Wees}}
\affiliation[University of Groningen]
{Physics of Nanodevices, Zernike Institute for Advanced Materials, University of Groningen, Nijenborgh 4, Groningen, AG 9747, The Netherlands}

\title{Magnon confinement in an all-on-chip YIG cavity resonator using hybrid YIG/Py magnon barriers}


\begin{document}

\begin{tocentry}

\includegraphics[width=8.25cm]{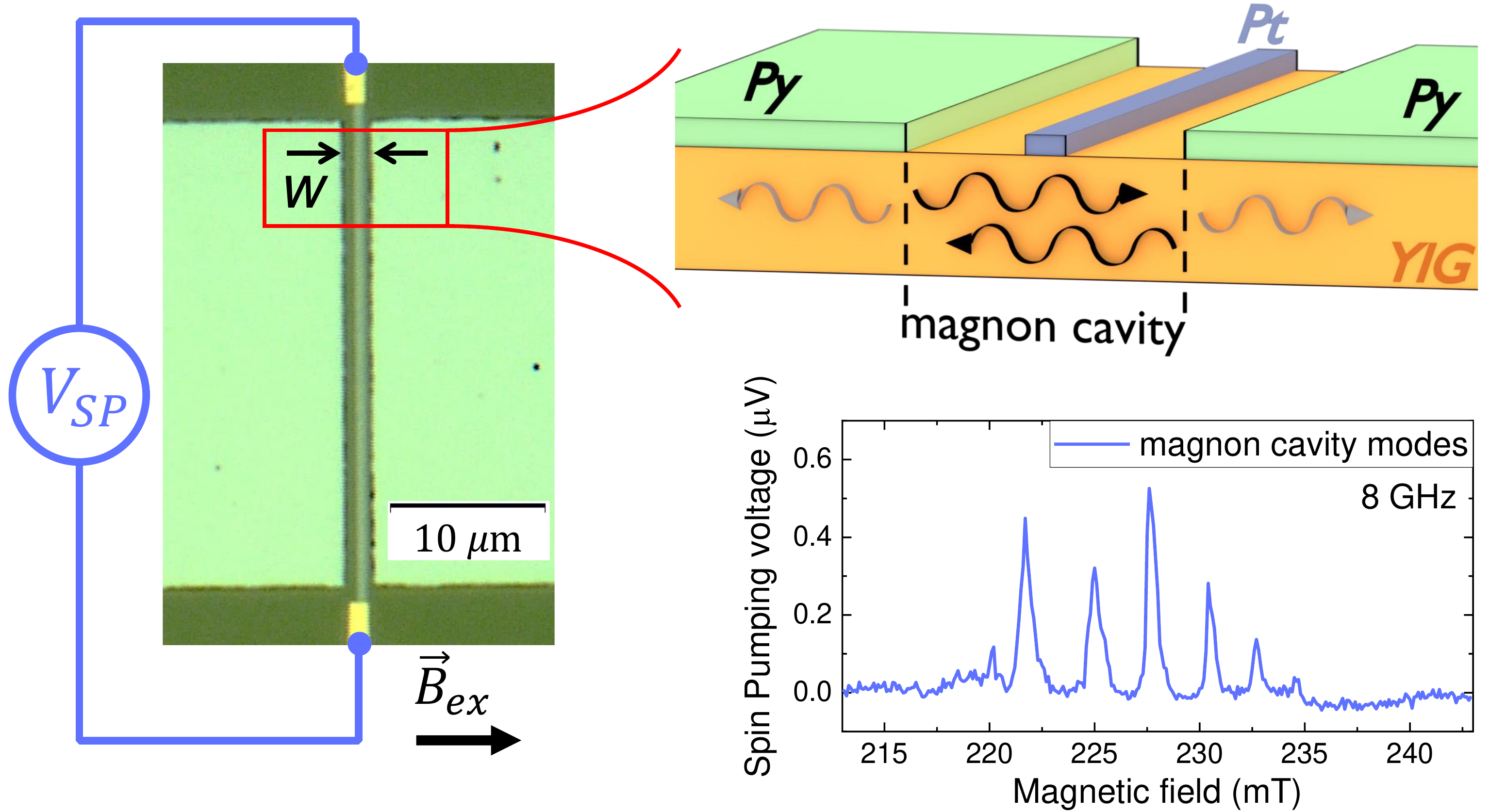}

\end{tocentry}

\begin{abstract}
Confining magnons in cavities can introduce new functionalities to magnonic devices, enabling future magnonic structures to emulate established photonic and electronic components.
As a proof-of-concept, we report magnon confinement in a lithographically defined all-on-chip YIG cavity created between two YIG/Permalloy bilayers.
We take advantage of the modified magnetic properties of covered/uncovered YIG film to define on-chip distinct regions with boundaries capable of confining magnons.
We confirm this by measuring multiple spin pumping voltage peaks in a 400 nm wide platinum strip placed along the center of the cavity.
These peaks coincide with multiple spin-wave resonance modes calculated for a YIG slab with the corresponding geometry.
The fabrication of micrometer-sized YIG cavities following this technique represents a new approach to control coherent magnons, while the spin pumping voltage in a nanometer-sized Pt strip demonstrates to be a non-invasive local detector of the magnon resonance intensity.

\end{abstract}

Magnonics aims to transmit, store, and process information in micro- and nano-scale by means of magnons, the quanta of spin waves.\cite{kruglyak2010magnonics,rezende2020fundamentals} 
Typical studies in this field use the ferrimagnet Yttrium Iron Garnet (Y\textsubscript{3}Fe\textsubscript{5}O\textsubscript{12} - YIG) as a key material, due to its desirable properties, such as very low magnetic losses,\cite{rezende2020fundamentals}
large applicability in mainstream electronics,\cite{pozar2011microwave}
long magnon spin relaxation time,\cite{Cornelissen2015}
and high magnon spin conductivity.\cite{Wei2022,Jungfleisch2022}
The magnetic proprieties of YIG
can be modified by the presence of strong exchange/dipolar coupling caused by the deposition of a ferromagnetic layer onto the YIG film.\cite{grunberg1981magnetostatic,li2020coherent}
These hybrid magnonic systems, such as YIG/Py,\cite{das2020modulation,fan2020manipulation,li2020coherent,Xiong2020,santos2021efficient,li2021phase} YIG/CoFeB,\cite{cramer2019orientation,qin2018exchange} and YIG/Co,\cite{klingler2018spin,wang2021nonreciprocal} draw attention from fundamental and applied physics towards the control of coherent magnon excitations for information processing.\cite{li2020hybrid}   
Alternatively, systems in which the magnetic material strongly interacts with electrodynamic cavities also provide a good platform for next-generation quantum information technologies, using the dual magnon-photon nature to enable new quantum functionalities.\cite{elyasi2020resources,harder2021coherent}
The field called cavity magnonics, among many applications, can offer good compatibility with CMOS, room temperature operation, and GHz-to-THz spin-wave transducers.\cite{barman20212021,pirro2021advances,sheng2021magnonics,chumak2022roadmap}

The possibility of confining magnons in cavities may enable future magnonic devices that emulate established electronic and photonic components, such as magnonic quantum point contacts, magnonic crystals, magnonic quantum bits, magnonic frequency combs, among others.\cite{Zakeri2020Magnonic,Yu2020Magnon,rameshti2021cavity,Hong2021,qin2021nanoscale,Vogel2015Optically,Wang2021, Hula2022,YUAN2021Quantum}
Building upon similar ideas, recent theoretical results suggest all-on-chip structures to produce magnonic cavities by magneto-dipole interaction with a chiral magnonic element,\cite{Kruglyak2021}
using an array of antiferromagnets on a ferromagnet,\cite{Xing2021}
or by proximity with superconductors.  
\cite{yu2022efficient}

One approach to confining magnons involves the fabrication of rectangular or circular nano- and micro-structures using YIG.\cite{trempler2020integration,costa2021compact,Lee-Wong2020,Srivastava2023}
Many of these structures are made by etching the YIG film or sputtering YIG from a target on defined micro-structures, adding an extra step in the fabrication process and limiting the options of available structures.
Recently,  Qin \textit{et al.}, demonstrated the (partial) confinement of magnons in a region of the YIG film covered by a ferromagnetic metal.\cite{qin2021nanoscale}
This confinement arises from the difference in magnetic properties between the covered and uncovered regions of the YIG film, resulting in magnon reflections at the boundaries.
Such phenomena enabled the fabrication of an on-chip nanoscale magnonic Fabry-P\'erot cavity,\cite{qin2021nanoscale, Hong2021} observed by the transmission of magnons through the YIG film.\cite{Vogel2015Optically}


\begin{figure}
\includegraphics[width=0.45\textwidth]{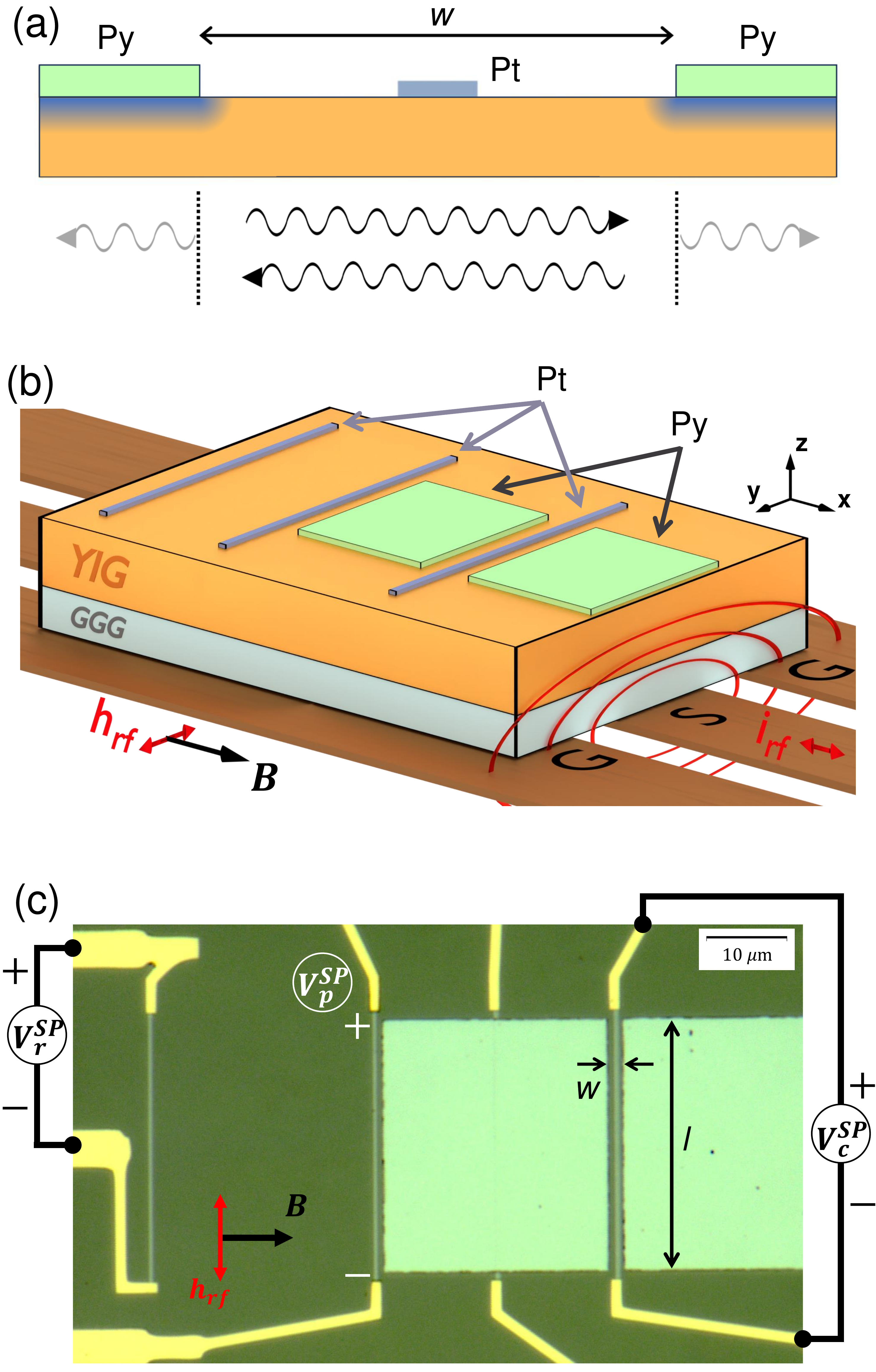}
\caption{\label{fig:skech}
(a) Schematic illustration of the lateral view of the cavity formed by confining the YIG film between two YIG/Py bilayers. 
The Pt strip in the middle of the cavity can be used as a local magnon detector.
(b) Schematic illustration of the waveguide stripline and sample arrangement of the microwave excitation setup for the FMR absorption and spin pumping measurements.
(c) Optical image of the fabricated devices, and the electrical connections used in the experiments.
The cavity is formed in the YIG region constrained between the Py squares, defined by $w\times l$. 
Multiple peaks are observed with the platinum strip placed along the center of the cavity, $V_{c}^{SP}$, and absent in both Pt strips placed outside of the cavity, $V_{r}^{SP}$ and $V_{p}^{SP}$.
}
\end{figure}

In this letter, we complement these studies by achieving an order of magnitude higher magnon confinement in a YIG film region which is \textit{not covered} by the ferromagnetic metal,
preserving the optimal magnetic properties of YIG within the cavity.
We report, as a proof-of-concept, the fabrication of an all-on-chip micrometer-sized YIG cavity, by partially covering 100 nm thick LPE-grown YIG film with two square-shaped 30 nm thick permalloy (Py) layers.
The exchange/dipolar interactions in the YIG/Py bilayer define on-chip, magnetically distinct regions, effectively forming reflecting boundaries for magnons, resulting in a magnonic cavity.
We confirm this by measuring multiple spin pumping voltage $(V^{SP})$ peaks with a 400 nm wide Pt strip placed along the center of the cavity.
This voltage is proportional to the intensity of the FMR-excited magnon resonances in the \textit{uncovered} YIG film, indicating the formation of standing-wave resonance modes.
We assign these peaks to multiple standing backward volume spin wave modes (BVSWs) and magnetostatic surface spin wave modes (MSSWs), calculated from the spin-wave theory for a YIG slab with similar dimensions. 
Multiple spin pumping voltage peaks were not observed for Pt strips placed outside of the cavity.
All the measurements were performed at room temperature.

\begin{figure}
\includegraphics[width=0.45\textwidth]{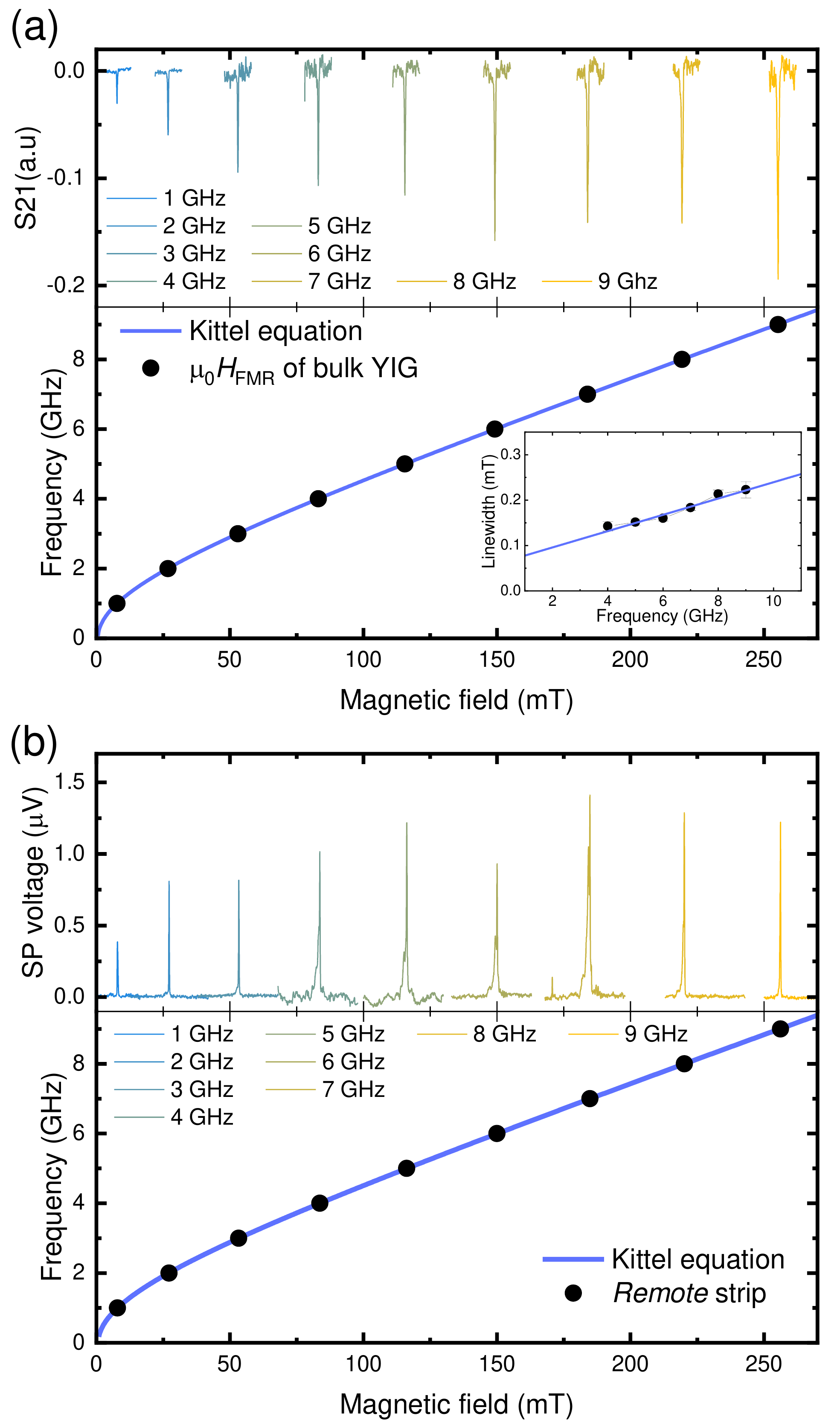}
\caption{\label{fig:FMR_SP}
Comparison between FMR absorption of the bulk YIG and the spin pumping voltage in the \textit{remote} strip. 
(a) Top panel shows $B$-field scan of the microwave transmission absorption, S21. The bottom panel shows the field position of FMR peaks for different microwave frequencies. The inset shows the linewidth \textit{vs.} resonance frequency.
Note that we do not see the magnon spectra because the magnetic field is uniform on a relatively long scale.
Therefore, only the uniform mode is excited and measured.
(b) The top panel shows the $B$-field scans of the $V^{SP}$ of the \textit{remote} Pt strip for different microwave frequencies.
The bottom panel shows the field position of the maximum $V_r^{SP}$ as a function of the microwave frequency.}
\end{figure}

The presence of BVSWs and MSSWs modes in micrometer YIG cavities has already been observed.\cite{heyroth2019monocrystalline,zhu2020waveguide,trempler2020integration,costa2021compact,kok-wai1986Magnetostatic,Ishak1988Tunable}
However, the YIG cavities were produced by sputtering or wet-etching technique, and the modes were measured by a local FMR antenna or time-resolved magneto-optical Kerr effect.
To the best of our knowledge, this work represents the first observation of cavity resonant modes measured by spin pumping voltages using all-on-chip hybrid magnonic structures, illustrated in Figure \ref{fig:skech} a).
The device was fabricated on a high-quality 100 nm thick YIG film grown by liquid phase epitaxy on a GGG substrate.
Electron beam lithography was used to pattern the device, consisting of multiple strips of Pt and two Py squares with edge-to-edge distance of $w=2$ $\mu$m, Figure \ref{fig:skech} b) and c).
The square shape was chosen to avoid effects of shape anisotropy in the Py film.\cite{mruczkiewicz2017spin,talapatra2020linear}
The sample was placed on top of a stripline waveguide and connected to a vector network analyzer.
The stripline waveguide was then placed between two poles of an electromagnet in such a way that the DC external magnetic field, $B=\mu_0H$, where $\mu_0$ is the vacuum permeability, and the microwave field, $h_{rf}$, were perpendicular to each other and both were in the plane of the YIG film in all the measurements, see Figure \ref{fig:skech} a) and b).
See supporting information section I for more details on sample fabrication and experimental setup.\cite{Supp}

The $B$-field scan of the microwave absorption, S21, which measures the overall magnetic response of $4\times3$ mm sized YIG film, is shown in the top panel of Figure \ref{fig:FMR_SP} a), for 1 to 9 GHz.
The FMR absorption peak of the 100 nm thick YIG film has a typical linewidth of $\approx0.2$ mT, demonstrating the high-quality of the YIG film.
The FMR spectra are fit using an asymmetric Lorentizan function, obtaining the $B$-field value of the FMR peak and the linewidth (FWHM), see details in Supporting Information section I.
The bottom panel of Figure \ref{fig:FMR_SP} a) shows the microwave frequency versus the $B$-field value of the FMR peak.
The solid blue curve corresponds to the best fit to the Kittel equation, $f=\gamma\mu_0/2\pi \sqrt{H(H+M)}$,\cite{pozar2011microwave}
where $\gamma$ is the gyromagnetic ratio.
The best fit was obtained for $\gamma/2\pi=27.2\pm0.1$ GHz/T and $M=142.4\pm0.8$ kA/m.
The inset in Figure \ref{fig:FMR_SP} a) shows the linewidth as a function of the microwave frequency.
The solid blue curve corresponds to the best linear fit, obtaining a Gilbert damping of $\alpha \approx 5.0\times10^{-4}$ and the inhomogeneous linewidth of ${\mu_0\Delta H}_0= 0.06$ mT.
In this letter, we address the uniform FMR resonance of the YIG film as "bulk" YIG resonance, or $\mu_0H_{FMR}$, to distinguish the FMR absorption measurement of the mm-range size YIG film from the local spin pumping voltage measurements for different platinum strips.
The upper panel of Figure \ref{fig:FMR_SP} b) shows the $B$-field scan of the spin-pumping voltage for the \textit{remote} Pt strip, $V_r^{SP}$, for different microwave frequencies.
The bottom panel presents the magnetic field of the $V_r^{SP}$ peak for different \textit{rf} frequencies.
Again, we fit the results using the Kittel equation with $\gamma/2\pi=27.2\pm0.1$ GHz/T and $M=140.2\pm0.9$ kA/m.

It is important to emphasize that albeit Figures \ref{fig:FMR_SP} a) and b) look effectively identical, they correspond to two completely different experiments.
Both measure the intensity of the ferromagnetic resonance, but in one case we measure the FMR absorption of the bulk YIG film, and in the other we measure the local spin pumping voltage, as a result of the injection of spin current by means of the spin pumping effect,\cite{tserkovnyak2002enhanced} and the conversion of the spin current into charge current in the Pt strip by the inverse spin Hall effect.\cite{azevedo2011spin,sinova2015spin}
We did not observe a significant peak broadening caused by the spin absorption due to the presence of the Pt layer.
This is because the width of the strip is 400 nm, such that it covers only a fraction of the YIG film resulting in a spin pumping response proportional to the FMR absorption of the bulk YIG.\cite{cheng2020nonlocal}
These results show that the platinum strip is a local and non-invasive intensity detector of the magnon excitation.

\begin{figure}
\includegraphics[width=1.00\textwidth]{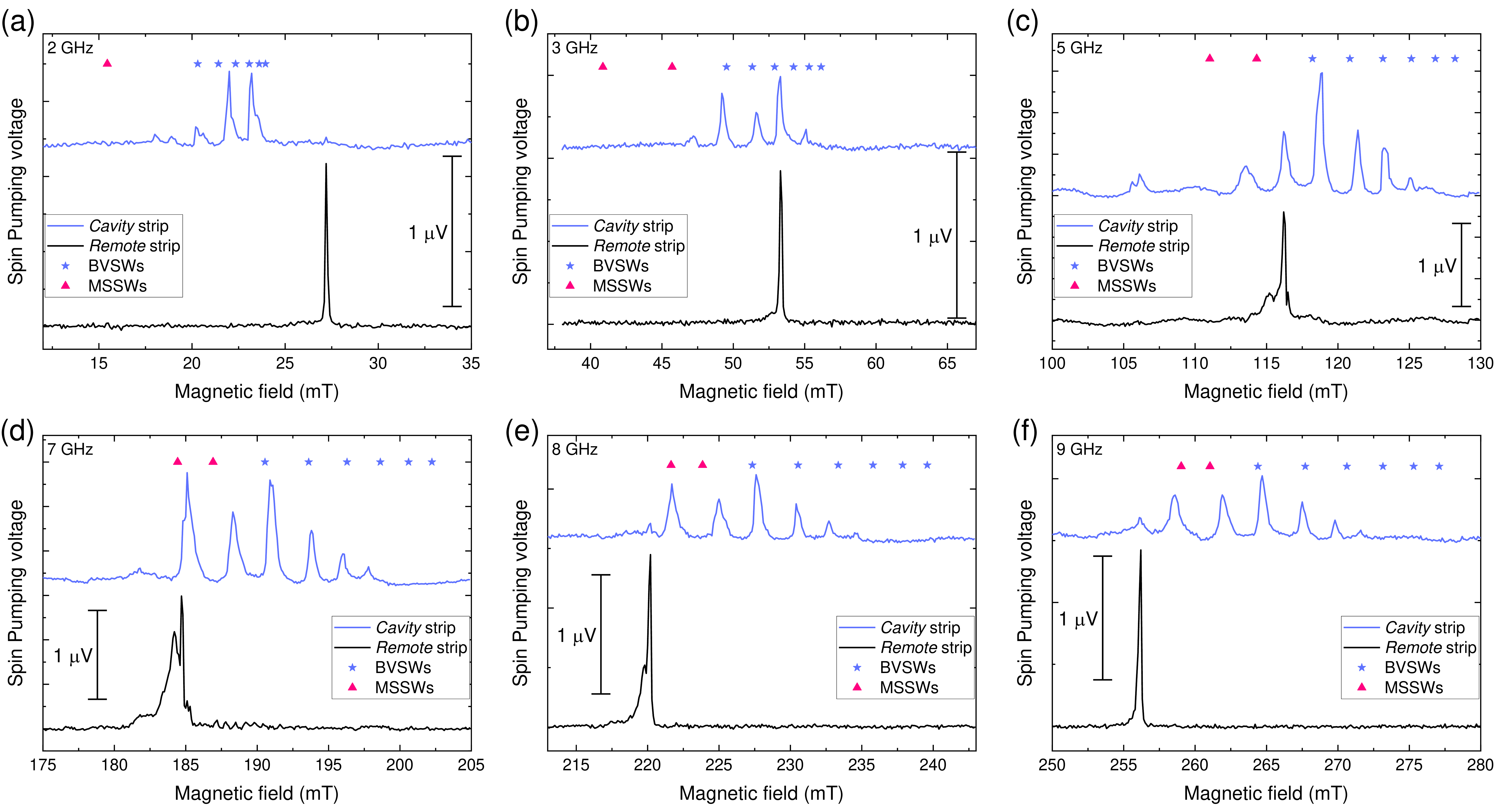}
\caption{\label{fig:different_frequencies}
The $B$-field scan of the spin pumping voltage measured with the \textit{remote} strip $V_r^{SP}$ and \textit{cavity} strip $V_c^{SP}$ for different frequencies is shown from (a) to (f).
The resonance modes of the \textit{cavity} strip occur before the FMR of the bulk YIG resonance field for low frequency, 2 GHz, and are present after the FMR bulk resonance for 9 GHz.
The resonance frequencies of the BVSWs and MSSWs modes obtained using equations \ref{eq:f_modes} and \ref{eq:f_factor} is shown as blue and pink star-symbol, respectively, for each frequency.
One can notice a secondary broad peak in the \textit{remote} strip at 7 GHz.
This peak is less evident or absent in other \textit{remote} strip.
We discuss more on that in the supporting information section III.\cite{Supp}
}
\end{figure}


Figure \ref{fig:different_frequencies} compares the $B$-field scan of the spin pumping voltage measured on the \textit{remote} strip, $V^{SP}_r$, solid black lines, and on the \textit{cavity} strip, $V^{SP}_c$, solid blue lines, for different \textit{rf} frequencies, from 2 GHz to 9 GHz.
The \textit{remote} strip shows a single resonance peak, while the \textit{cavity} strip shows multiple resonances. 
The average linewidth of the spin pumping voltage peaks on the \textit{cavity} strip is slightly broader than the \textit{remote} strip, suggesting an additional damping contribution.
Note that there is no pronounced peak on the \textit{cavity} strip $B$-field scan corresponding to the bulk YIG resonance at 2 GHz, and the corresponding peak is small for 8 and 9 GHz, Figure \ref{fig:different_frequencies} a), e) and f), respectively.
This indicates that the spin pumping voltage on the \textit{cavity} strip is dominated by magnons excited inside the cavity itself, not by magnons generated outside of the cavity, corresponding to bulk FMR values, which could be transmitted into the cavity.
As mentioned above, $V^{SP}_c$ is proportional to the resonance intensity of the YIG film in the region between the Py squares.
This means that the series of resonance modes present underneath the Pt strip, indicate the existence of a cavity supporting standing magnonic waves.


We can explain these modes by calculating the spin-wave dispersion relation for a YIG slab with dimensions $w\times l$ with magnetization in the plane of the film, given by\cite{guslienko2005boundary,mahmoud2020introduction,Zheng2022Spin}
\begin{equation}
f = \frac{\gamma\mu_0}{2\pi}{\Big(\left(H+H_{a}+\lambda_{ex}k_{tot}^2 M\right)
\left(H+H_{a}+\lambda_{ex}k_{tot}^2 M+M\mathbb{F}\right)\Big)}^{1/2},
\label{eq:f_modes}
\end{equation}
where $\lambda_{ex}=2A/\mu_0{M}^2$ is the exchange constant with $A=3.5$ pJ/m, ${k_{tot}}^2={k_n}^2+{k_m}^2$ is the total quantized wavenumber defined by $k_n=n\pi/w$ and $k_m=m\pi/l$, where $n$ and $m$ are the mode numbers along the width and length of the cavity, respectively. The function $\mathbb{F}$ can be written as
\begin{equation}
\mathbb{F} = \mathbb{P}+\bigg(1-\mathbb{P}\left(1+\textrm{cos}^2\left(\phi_k-\phi_M\right)\right)+\frac{M\mathbb{P}\left(1-\mathbb{P}\right)\textrm{sin}^2\left(\phi_k-\phi_M\right)}{H+M\lambda_{ex}{k_{tot}}^2}\bigg),
\label{eq:f_factor}
\end{equation}
where $\mathbb{P}=1-\frac{1-e^{dk_{tot}}}{dk_{tot}}$, for a YIG film with thickness $d$.
In Eq. \ref{eq:f_factor}, $\phi_k=\textrm{arctan}\left(k_m/k_n\right)$, and $\phi_M$ is the angle between the magnetization and the spin-wave propagation or direction of $\vec{k}$.
Two magnetostatic modes can be accessed considering the symmetry of the device, the magnetostatic surface spin wave modes (MSSWs), where $\vec{k}\bot \vec{M}$, \textit{i.e.}, $\phi_M=\pi/2$ and the backward volume spin waves modes (BVSWs) where $\vec{k}\parallel \vec{M}$, \textit{i.e.}, $\phi_M=0$.

The best fit to Eq. \ref{eq:f_modes} and \ref{eq:f_factor} reproducing the majority of the spin pumping peaks for different frequencies was obtained using $M=130$ kA/m, $d=100$ nm, $w=2.5$ $\mu$m, $l=30$ $\mu$m,
$\mu_0H_a=10$ mT,
and $\gamma/2\pi=26.5$ GHz/T.
The calculated values for $(m=1)$ of the $n=(1,2,3,...6)$ mode of the BVSWs and first and second mode of MSSWs are shown in Figure \ref{fig:different_frequencies} a) to f), blue, and pink star symbols, respectively.
We obtained a good agreement between the peaks present in $V^{SP}_c$ and the calculated modes.
Since $l\gg w$, modes with $m>1$ are hard to distinguish since they are superimposed on the $m=1$ mode due to their proximity in frequency.
One important feature obtained from the fit is an anisotropy field of $\mu_0H_a=10$ mT.
The anisotropy may originate from the stray fields produced by the magnetization of Py, leading to a local increase of the effective DC-magnetic field applied on the cavity.\cite{Zhang2019Controlled}
Py stray fields may also be responsible for inducing even modes in the cavity, observed in our results.\cite{Zhang2019Controlled}
These even modes are not expected for a YIG slab when a homogeneous DC magnetic field and \textit{rf} field are applied.\cite{trempler2020integration,costa2021compact}

\begin{figure}
\includegraphics[width=0.5\textwidth]{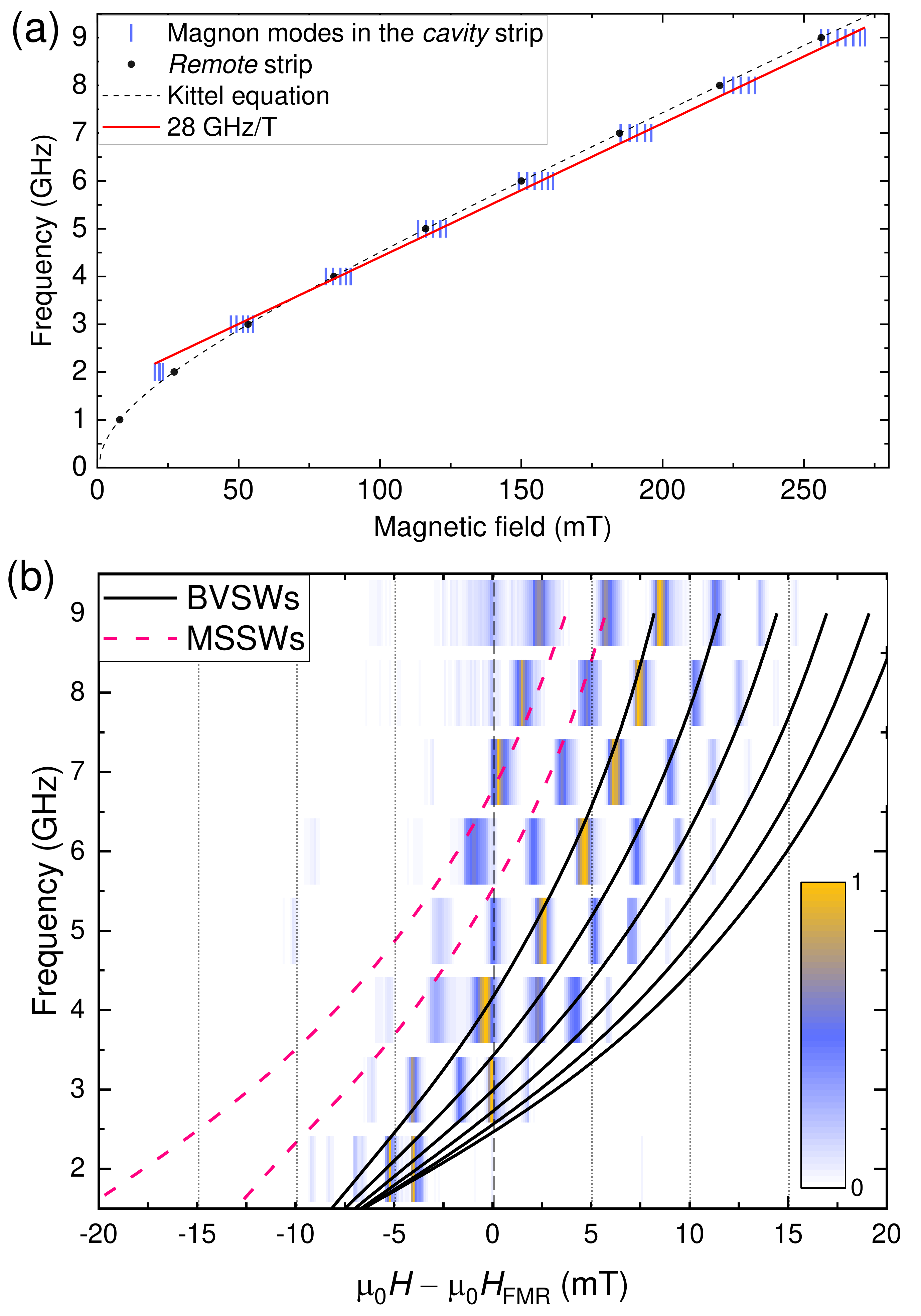}
\caption{\label{fig:SP_spectra}
(a) Distribution of the identified magnon modes in the cavity as a function of the magnetic field for each frequency, plotted as vertical blue stripes. The solid red curve corresponds to $28$ GHz/T.
The resonance of the \textit{remote} strip and the best fit of the Kittel equation are addressed as a black symbol and a dashed black line, respectively.
(b) Spin pumping intensity spectra of the Pt strip within the cavity.
The solid black and dashed pink lines correspond to the resonant modes for a YIG slab with similar dimensions as the build cavity. 
}
\end{figure}

One can see that the modes appear in Figure \ref{fig:different_frequencies} at a lower field than the bulk YIG resonance for 2 GHz and a higher field than the bulk resonance at 9 GHz.
We can analyze the frequencies of the cavity resonance modes by simultaneously plotting the SP-FMR resonance field of the \textit{remote} strip (black circles) and the resonance field of each magnon mode in the \textit{cavity} strip (vertical blue stripes), Figure \ref{fig:SP_spectra} a).
Overall, the resonance field distribution of cavity modes evolves linearly in frequency, with a slope of $\approx 28$ GHz/T, solid red curve in Figure \ref{fig:SP_spectra} a).
The linear behavior was also reported in previous YIG cavity results.\cite{dai2020octave,costa2021compact,kok-wai1986Magnetostatic}

To emphasize the localized magnon detection characteristic of the Pt strip, we normalize the spin pumping voltage of the \textit{cavity} strip, $V_c^{SP}$, between 0 and 1, based on the lowest and highest spin pumping voltage value in each $B$-field scan.
We centered the $B$-field scans with respect to the resonance field of the bulk YIG obtained from the FMR measurements, $\mu_0H_{\textsc{FMR}}$, for each frequency.
Figure \ref{fig:SP_spectra} b) shows the dispersion of the resonant modes of the \textit{cavity} strip compared with FMR of the bulk YIG.
The BVSWs modes, up to $(n=6)$, and the first and second modes of the MSSWs calculated from Eq. \ref{eq:f_modes} and \ref{eq:f_factor} are shown in Figure \ref{fig:SP_spectra} b) as solid black and dashed pink lines, respectively.
The model of Eq. \ref{eq:f_modes} and \ref{eq:f_factor} is very useful for confirming the mode position as a function of field and frequencies, and the peak spacing between each peak.
Presenting the voltages as intensity spectra demonstrates that the 400 nm wide Pt strip can be used as a localized FMR detector in future magnonic cavity studies.
The spectrum at 1 GHz is not shown in Fig. \ref{fig:SP_spectra} b) due to the absence of a prominent spin-pumping voltage peak.
Additional intensity spectra with other cavity widths are shown in the supporting information section II,\cite{Supp} confirming the reproducibility of the fabrication technique.

We do not observe multiple voltage peaks with the \textit{proximity} strip, $V_p^{SP}$, placed close to the Py square but still outside of the cavity.
This confirms that the Pt layer alone does not induce sufficient magnetic changes in the YIG to create a cavity.
Additionally, we do not observe multiple peaks in a device where the Py film was replaced with gold, ruling out microwave artifacts and confirming the requirement of confinement by YIG/Py bilayers on both sides to create the cavity, see supporting information section III.\cite{Supp}
We hypothesize that the difference between the magnetization dynamics of covered and \textit{uncovered} YIG regions by the Py film creates a magnon barrier, as ilustrated in Figure \ref{fig:skech} a), and discussed in previous reports.\cite{Hong2021,qin2021nanoscale}

\begin{figure}[t]
\includegraphics[width=1.0\textwidth]{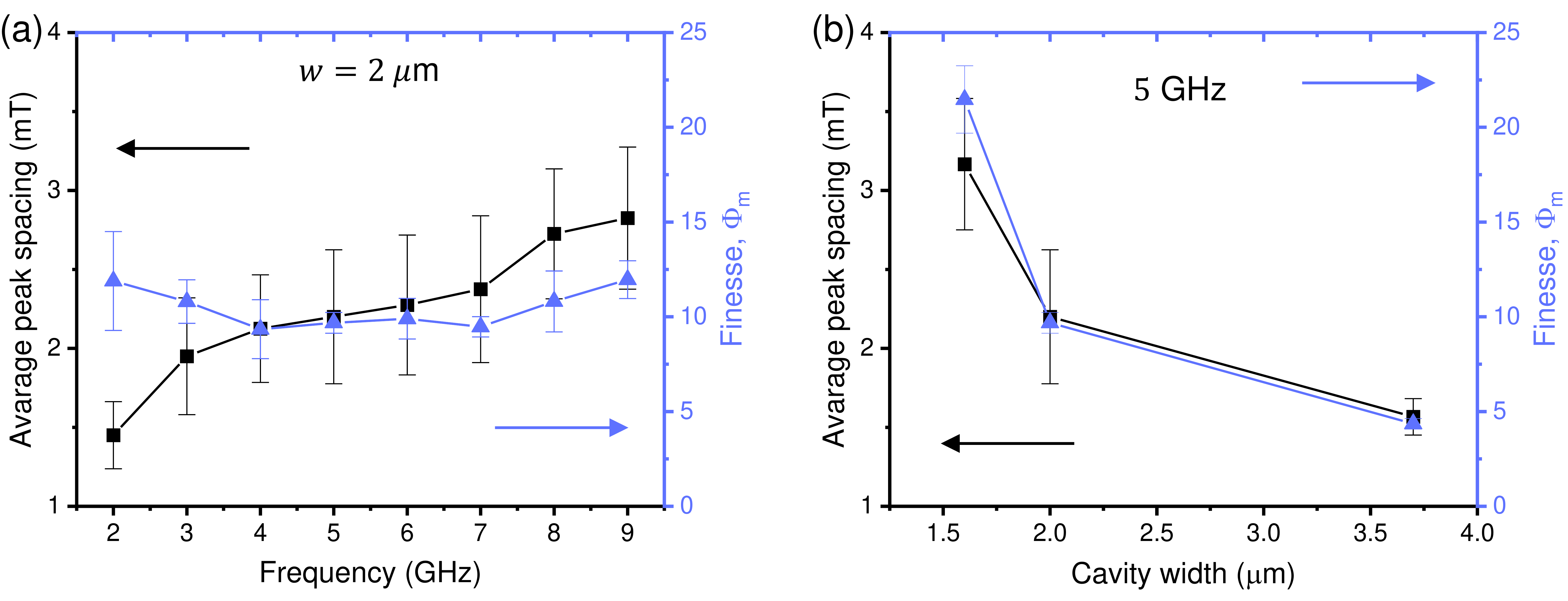}
\caption{\label{fig:sketch_cavity}
(a) Frequency dependence of the average $\Delta B_{spacing}$ and the finesse $\Phi_m$. The error bar is calculated as the standard deviation of $\Delta B_{spacing}$.
(b) Average peak spacing and finesse as a function of cavity width calculated from the $B$-field scan at 5 GHz.}
\end{figure}

In analogy to a (lossy) cavity resonator, we can estimate a finesse given by $\Phi_m=\Delta B_{spacing}/\Delta B$, where $\Delta B_{spacing}$ is the $B$-field peak spacing and $\Delta B$ is the linewidth of the resonance peak.\cite{Ismail2016Fabry}
As a figure of merit, the frequency dependence of the average $\Delta B_{spacing}$ and calculated finesse is shown in Figure \ref{fig:sketch_cavity} a).
Although the peak spacing increases as a function of frequency, the finesse fluctuates around a value of $\Phi_m\approx 10$, about one order of magnitude higher than the previous report.\cite{qin2021nanoscale}
This value of finesse corresponds to a reflectance of $R\approx 0.73$, approximately constant through the entire frequency range.
Figure \ref{fig:sketch_cavity} b) shows the average peak spacing and the finesse as a function of the cavity width, calculated from the $B$-field scans at 5 GHz, see supporting information section II.\cite{Supp} 
Although only three sets of data are presented, a clear correlation between the average peak spacing and the finesse is observed.
This correlation arises because the cavity is formed in a \textit{uncovered} YIG region, preserving the optimal magnetic properties of YIG.
In fact, the peak linewidth decrease with decreasing cavity width, $w$, indicating that the additional broadening may stem from the superposition of higher modes ($m>1$) along the cavity length, see supporting information section II.\cite{Supp}
The frequency spacing between these modes increases as a function of the aspect ratio $(l/w)$.
Ultimately, a maximum finesse of $\Phi_m\approx 21$, or $R\approx0.86$, is achieved for a cavity with $w=1.6$ $\mu$m, which demonstrates a high potential for magnon confinement.

We consider Eq. \ref{eq:f_modes} and \ref{eq:f_factor} as an approximate model since it describes a YIG slab with dimensions $w\times l$, where the magnetization amplitude is minimum at the boundary, \textit{i.e.}, an infinite potential well.
In our case, the cavity is a consequence of the exchange and dipolar interaction in the YIG/Py bilayer.\cite{kalinikos1986theory,grunberg1981magnetostatic,li2020coherent,qin2021nanoscale}
This means that a finite height potential well would better describe the system.
This discrepancy can be the origin of a minor deviation between the measured $V_c^{SP}$ peaks and the calculated cavity modes.
Accurate modeling of the modes should be performed using micromagnetic simulations, taking into account the exchange and dipolar interaction with Py in further investigation.
The potential height of the cavity barriers should be dependent on the thickness of the YIG film and the exchange/dipolar interaction with the top ferromagnetic layer.
Further investigation using the present technique should be performed for different adjacent ferromagnetic layers in which strong exchange interaction has already been reported, such as YIG/CoFeB\cite{qin2018exchange} and YIG/Co.\cite{klingler2018spin}
YIG films thinner than 100 nm with the damping below than $5.0\times10^{-4}$ are good candidates to produce magnonic cavities with higher reflectance factors keeping the magnetic losses close to those reported in this letter.\cite{Schmidt2020Ultra}

In summary, we take advantage of the difference in the magnetic dynamics between the YIG film and the YIG/Py bilayers, to fabricate an all-on-chip magnonic cavity supporting standing magnon modes in a \textit{uncovered} YIG film between two YIG/Py bilayers.
This approach enables the confinement of magnons while preserving the optimal magnetic properties of the YIG cavity.
The spin pumping voltage of a 400 nm wide Pt strip proved to be a reliable technique to detect the magnon resonance modes of the cavity.
Following this idea, 1D and 2D magnonic crystals could be obtained by having a regular array of magnetic strips onto YIG, with the possibility of measuring the magnon modes locally by means of spin pumping.
Moreover, further investigations should involve designing coupled cavities by placing two cavities side by side, where the coupling strength could be controlled by the width of the central YIG/Py bilayer.
This cavity fabrication process opens new possibilities for investigating and characterizing micron-sized YIG cavities with a wide range of arbitrary shapes. 
It also allows for the implementation of on-chip magnonic computation structures, serving as a printed circuit board for magnons.
These results demonstrate a promising combination of hybrid magnonics and cavity magnonics, which has the potential to drive the integration of future all-on-chip magnonic devices into mainstream microwave electronics.

\begin{acknowledgement}

We acknowledge the technical support from J. G. Holstein, T. J. Schouten  and H. de Vries, F. A. van Zwol, A. Joshua.
We are grateful to A. Azevedo, C. Ciccarelli and C. M. Gilardoni for the valuable discussion.
We acknowledge the financial support of the Zernike Institute for Advanced Materials and the Future and Emerging Technologies (FET) programme within the Seventh Framework Programme for Research of the European Commission, under FET-Open Grant No. 618083(CNTQC).
This project is also financed by the NWO Spinoza prize awarded to Prof. B. J. van Wees by the NWO, and ERC Advanced Grant 2DMAGSPIN (Grant agreement No. 101053054).

\end{acknowledgement}

\begin{suppinfo}

Device fabrication and experimental setup details (section I). Additional cavities measurements (Section II) and control samples measurements (section III).

\end{suppinfo}

\bibliography{achemso-demo}

\end{document}


\renewcommand\thefigure{S\arabic{figure}}
\renewcommand\theequation{S\arabic{equation}}    

\title{SUPPORTING INFORMATION \\
Magnon confinement in an all-on-chip YIG cavity resonator using hybrid YIG/Py magnon barriers}

\author{Obed \textbf{Alves Santos}}
\altaffiliation{\textbf{Correspondence should be addressed: \\}\href{mailto:oa330@cam.ac.uk}{oa330@cam.ac.uk}, or \href{mailto:obed.alves.santos@gmail.com}{obed.alves.santos@gmail.com}}
\affiliation{Physics of Nanodevices, Zernike Institute for Advanced Materials, University of Groningen, Nijenborgh 4, Groningen, AG 9747, The Netherlands}
\affiliation{Cavendish Laboratory, University of Cambridge, Cambridge, CB3 0HE, United Kingdom}
\author{Bart J. \textbf{van Wees}}%
\affiliation{Physics of Nanodevices, Zernike Institute for Advanced Materials, University of Groningen, Nijenborgh 4, Groningen, AG 9747, The Netherlands}

\maketitle

\vspace{-1cm}

\section{Sample fabrication and experimental setup}

\begin{figure}[b]
\includegraphics[width=0.9\textwidth]{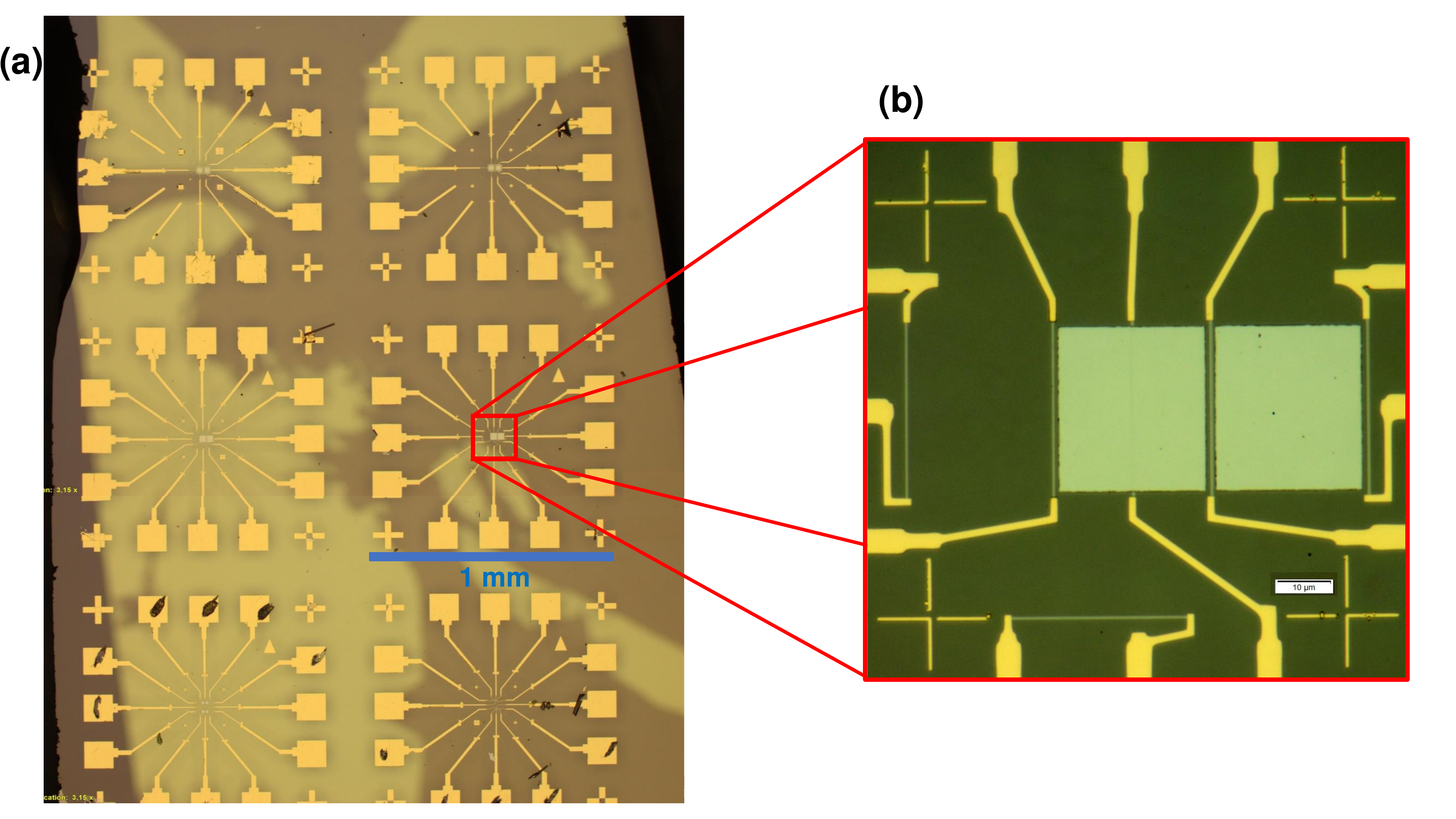}
\caption{\label{fig:wholesample}
\textbf{a)} Show the optical image of the final YIG sample with multiple devices. \textbf{b)} Show a zoom-in optical image of an individual device.
}
\end{figure}

The samples consists of a high-quality 100 nm thick YIG film $(\textrm{Y}_3\textrm{Fe}_5\textrm{O}_{12})$ grown by liquid phase epitaxy on a GGG substrate, obtained commercially from Matesy GmbH, measuring approximately $(4\times3)$ mm.
Electron beam lithography (EBL) was used to pattern the device, which consists of multiple strips of Pt with 35 $\mu$m length and 400 nm wide.
The Permalloy (Py) squares have dimensions $(30\times30\;{\mu\textrm{m}}^{2})$, the square shape was chosen to avoid shape anisotropies of the Py film.\cite{mruczkiewicz2017spin,talapatra2020linear}
The Pt and Py layers were deposited by DC sputtering in an Ar+ plasma with thicknesses of 8 nm and 30 nm, respectively.
The deposition of Ti(5 nm)/Au(75 nm) leads is made by e-beam evaporation.
The last step consists of mounting the sample onto a non-resonant stripline waveguide, which can be used from 0.1 to 9 GHz with a characteristic impedance of 50 $\Omega$, 0.030" RO4350, GCPWG, with a signal line 45 mil (1.14 mm) wide.
Figure \ref{fig:wholesample} a) show an optical image of the final YIG film sample containing multiple devices, while b) shows a zoom-in image of an individual device.
The wire bonding is made using AlSi (Al 99\%, Si 1\%) wires on the sample holder and connected to a lock-in amplifier.

The microwave field excitation is sent connecting the stripline waveguide to a vector network analyzer (VNA).
The FMR absorption of the YIG film is obtained by scanning the absorption (S21) or reflection (S11) in the VNA for a fixed microwave frequency as a function of the magnetic field.
To obtain the $\mu_0H_{\textsc{FMR}}$ value and the FMR linewidth $(\Delta H)$, each S21 $B$-field scan is fitted using an asymmetric Lorentzian function,
$L(H-H_{FMR})=S\Delta H^2/[(H-H_{FMR})+\Delta H^2]+A[\Delta H(H-H_{FMR}]/[(H-H_{FMR})+\Delta H^2]$.
Where $S$ and $A$ are the symmetric and antisymmetric amplitudes.
The spin pumping voltage is measured at the end of the Pt strips by a lock-in amplifier, triggered with the VNA.
The microwave is then switched from low power $P_{rf}^{low}=-25$ dBm to high power $P_{rf}^{high}=16$ dBm, with a modulation frequency of $27.71$ Hz.
The (waveguide stripline + sample) is positioned between two poles of an electromagnet such that the external magnetic field $(H)$ and the microwave field $(h_{rf})$ are perpendicular to each other, and both are in applied in the plane of the YIG film.
All the measurements were performed at room temperature.

\begin{figure}[h]
\includegraphics[width=1.0\textwidth]{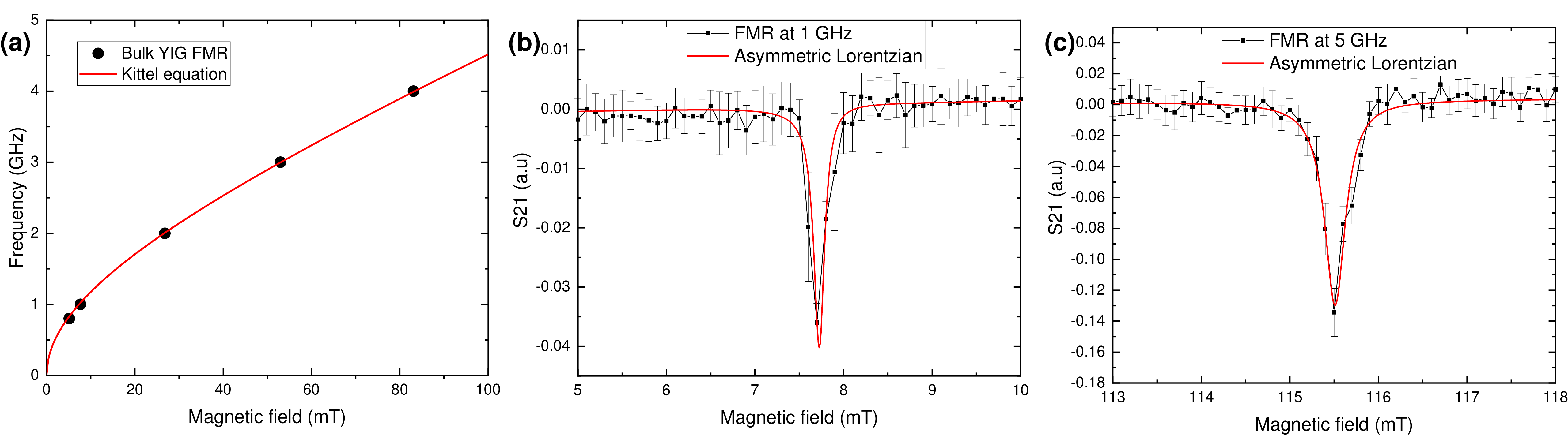}
\caption{\label{fig:rawFMRdata}
\textbf{a)} Field resonances of the bulk YIG FMR for low frequency.
\textbf{b)} and \textbf{c)} are S21 measurements of the bulk YIG FMR at 1 GHz and 5 GHz, respectively. Our external magnetic field step corresponds to 0.1 mT.
The error bar corresponds to the standard deviation from the multiple VNA readings. 
}
\end{figure}

In spintronics, the FMR drives the spin pumping effect (SPE),\cite{tserkovnyak2002enhanced} where a flow of spin current, occurs from the ferromagnetic/ferrimagnetic layer towards an adjacent layer at the peak of the microwave absorption during the FMR process.\cite{mizukami2001ferromagnetic}
Therefore, the YIG film injects a spin current by means of the SPE into the Pt strip.\cite{tserkovnyak2002enhanced}
That spin current can be expressed as
\begin{equation}
J_s=g_{eff}^{\uparrow\downarrow}\frac{\hbar\omega}{4\pi}
\left(
\frac{h_{rf}}{\Delta H}%
\right)^2
L(H-H_{FMR})
,
\label{eq:spin_current}
\end{equation}
where $g_{eff}^{\uparrow\downarrow}$ is the effective spin mixing conductance, $\omega=2\pi f$ is the $rf$ frequency, $\hbar$ is the reduced Planck constant, and $L(H-H_{FMR})$ is the FMR absorption, usually a Lorentizian-like line shape.
The spin current along $\hat{z}$ with spin polarization $\vec{\sigma}$ along $\hat{x}$ is converted into charge current along $\hat{y}$ direction by the inverse spin Hall effect (ISHE), following $\vec{J_c}\propto\theta_{Pt} (\vec{J_s}\times \vec{\sigma})$.\cite{sinova2015spin}
Where $\theta_{Pt}$ is the spin Hall angle, which quantifies the conversion efficiency between spin and charge currents.
The total voltage build-up at the edge of the Pt strip can be expressed by\cite{azevedo2011spin,sinova2015spin}
\begin{equation}
V^{SP}=\theta_{Pt}\frac{R_{Pt}\lambda_{Pt}w_{Pt}}{t_{Pt}}
\left(
\frac{2e}{\hbar}%
\right)
\textrm{tanh}\left(\frac{t_{Pt}}{2\lambda_{Pt}}\right)
J_s
,
\label{eq:SP_voltage}
\end{equation}
where $R_{Pt}$, $t_{Pt}$, $w_{Pt}$ and $\lambda_{Pt}$ are the resistance, thickness, width, and the spin diffusion length of the Pt strip.  

The spin pumping process typically leads to an increase in the FMR linewidth as it introduces an additional component to the magnetic losses of the ferromagnetic layer.
However, in our measurements, we did not observe any significant broadening of the linewidth in the spin pumping voltages measured on the Pt strip. 
This absence of pronounced linewidth broadening in $V^{SP}_r$ can be attributed to the fact that the Pt strip only covers a fraction of the YIG film.
Cheng \textit{et al.} demonstrated that for Pt strips with widths  $w<5$ $\mu$m, the spin pumping voltage is dominated by the spin current injected in the proximity of the Pt strip, within a YIG region unaffected by the presence of the Pt layer.
In this region, the damping and linewidth correspond to the bulk values. Hence, the spin pumping voltage measured with the Pt strip effectively acts as a localized FMR absorption detector or an rf antenna.\cite{cheng2020nonlocal}

One can realize that an additional spin pumping voltage could be detected by the Pt strip within the field range corresponding to the resonance of the Py layer, resulting from spin pumping from Py towards the YIG film. 
However, in our measurements, we observed that the voltage remained below the noise level throughout the resonance field of Py.
As a result, our main focus in this study is on the resonance of the YIG film.


\section{Additional cavities measurements}

\begin{figure}
\includegraphics[width=1.00\textwidth]{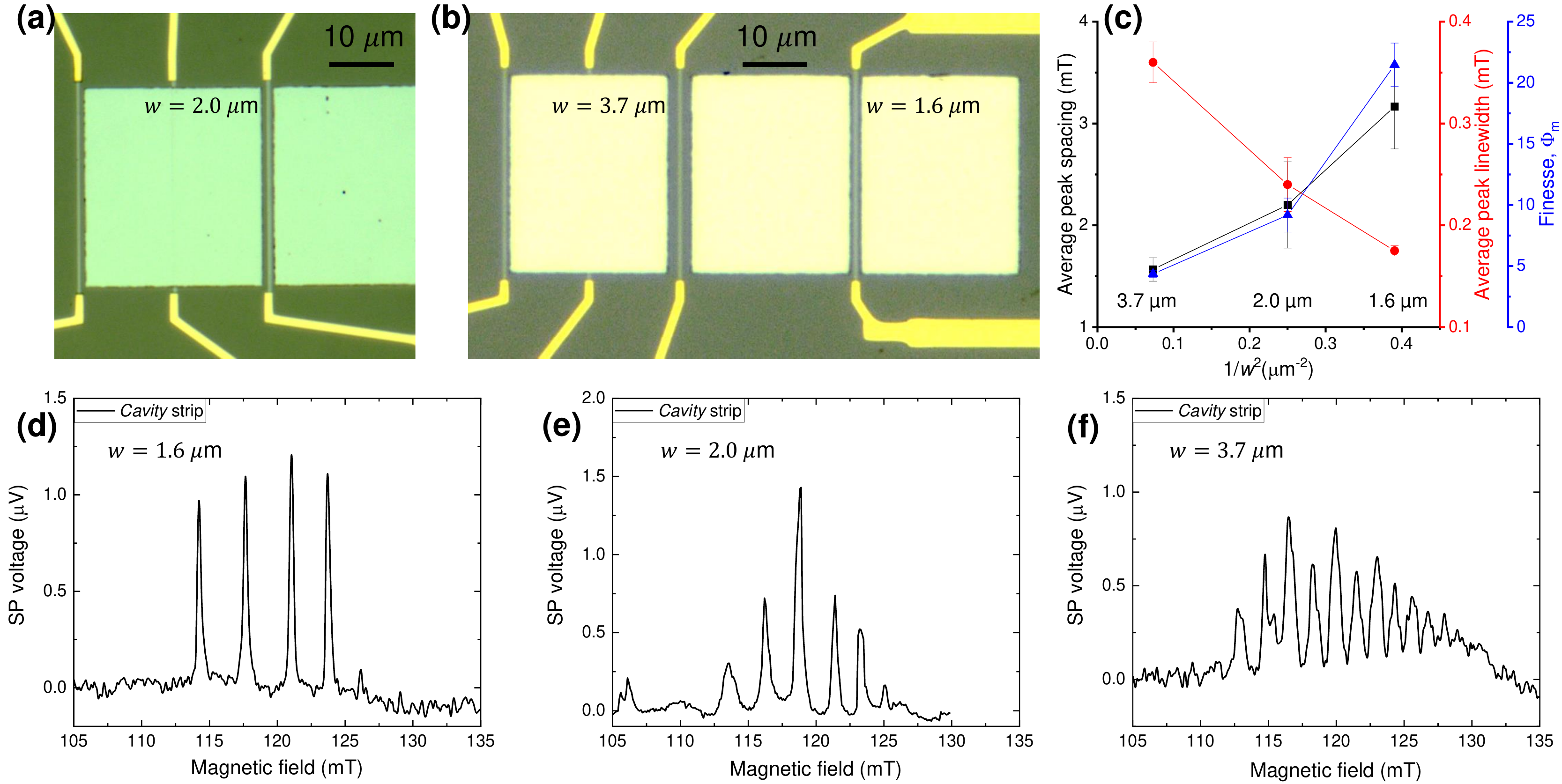}
\caption{\label{fig:diff_width}
\textbf{a)} and \textbf{b)} show optical images from different devices fabricated with distinct distance cavity widths, indicated in the figure.
\textbf{c)} Average peeks spacing in (mT), average peak linewidth, and calculated finesse as a function of the inverse square of the cavity width $(w)$.
\textbf{d)}, \textbf{e)} and \textbf{f)} shows $B$-field scan of the spin pumping voltage in the central platinum strip at 5 GHz, for a cavity width of $w=1.6$ $\mu$m, $w=2.0$ $\mu$m, and $w=3.7$ $\mu$m, respectively.
}
\end{figure}

\begin{figure}
\includegraphics[width=1.0\textwidth]{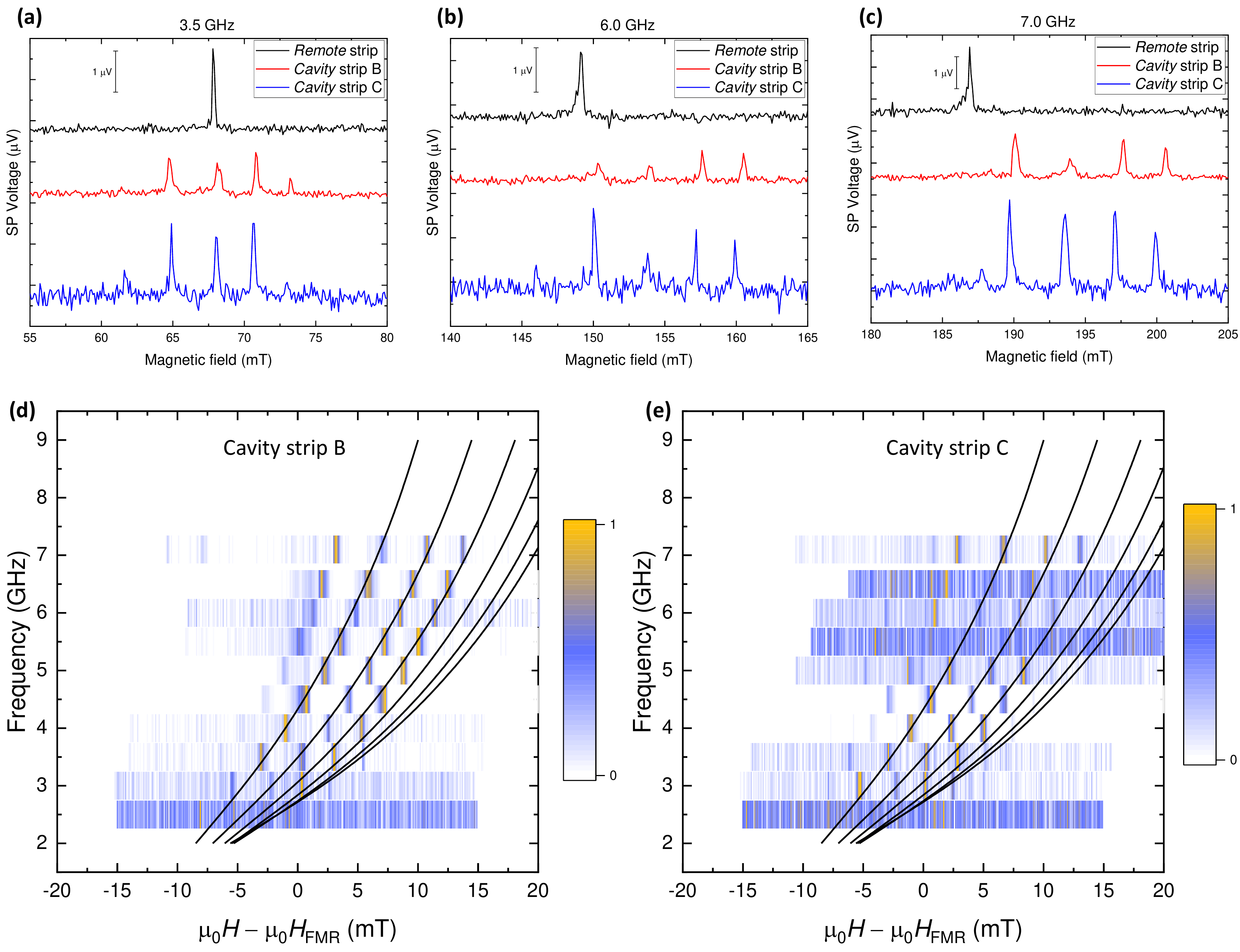}
\caption{\label{fig:Extra_cavities}
\textbf{a)} to \textbf{c)} shows the $B$-field scan of the spin pumping voltage obtained for the \textit{remote} strip and the Pt strips placed along the center of the cavity B and C, respectively. 
\textbf{d)} and \textbf{e)} shows the spin pumping intensity spectra for two different cavities with the same width distance of $w=1.6$ $\mu$m.}
\end{figure}

Additional spin pumping measurements were conducted on several other cavities to demonstrate the reproducibility of the method.
Optical microscope images of the cavities with different widths are shown in Figure \ref{fig:diff_width} a) and b).
The main results presented in the text are based on the cavity denoted as "Cavity A," which has a width of $w=2.0$ $\mu$m.
Subsequently, we present results for two separate cavities: "Cavity B" and "Cavity C," both with a width of $w=1.6$ $\mu$m, as well as "Cavity D", which has a width of $w=3.7$ $\mu$m.
Figure \ref{fig:diff_width} c) shows the average peak spacing, average peak linewidth, and finesse as a function of the inverse square of the cavity width.
It can be observed that the peak spacing shows a linear dependence as function of $1/w^2$ within the error bar, following the expected behavior for a magnonic cavity.\cite{yu2022efficient,Xing2021} 
The linear behavior is also evident in the average peak linewidth, suggesting that increasing the aspect ratio of the cavity ($l/w$) leads to a more homogeneous distribution of cavity modes.
However, it should be noted that our results are limited to only three different cavity widths, making it challenging to make a definitive statement about the observed behavior.
Figure \ref{fig:diff_width} d) to f) show the $B$-field scan of the spin pumping voltage for three cavities widths for 5 GHz.
As the width decreases, a few observations can be made.
Firstly, the peak linewidth decreases, indicating a narrower spectral distribution of the spin pumping voltage.
Secondly, the peak spacing increases, suggesting a larger separation in corresponding excitation frequencies.
Finally, the spin pumping voltage height becomes more equal for the cavity with $w=1.6$ $\mu$m.

Figure \ref{fig:Extra_cavities} a) to c), shows the $B$-field spin pumping voltage for the \textit{remote} cavity and cavities B and C.
All three measurements were obtained simultaneously, with each \textit{cavity} strip connected to an independent lock-in amplifier.
Both cavities, B and C, have been designed with the same width of $w= 1.6$ $\mu$m.
This explains why the field position of almost every peak aligns with each other.
Once again, it is evident that the corresponding peak for the bulk YIG resonance is absent or reduced to the noise level at 6 GHz and 7 GHz. 
This observation highlights that the voltage peaks are primarily determined by the magnon modes confined within the cavity.
One can also observe a small alternation of the intensity between consecutive peaks in Figure \ref{fig:Extra_cavities} a) to c).
A similar trend seams to be present in Figure 3 of the main text.
This behavior might occur due to the difference in microwave excitation corresponding to odd and even modes.
However, we cannot explain them in detail without more detailed micromagnetic simulations, taking into account the exchange and dipolar interactions in the YIG/Py bilayer, and then correctly addressing each peak to the corresponding magnon mode and intensity.

The spin pumping intensity spectra for cavities B and C are present in Figure \ref{fig:Extra_cavities} d) and e), with a frequency spacing of 0.5 GHz.
Although the peak between 5.5 GHz and 6.5 GHz in cavity C is difficult to identify due to the noise, both cavities exhibit a similar peak dispersion, with almost every peak matching at each frequency.
The solid black lines in Figure \ref{fig:Extra_cavities} d) and e) are the modes calculated by equation (1) and (2) from the main text, up to $n=6$ using the following parameters: $w=1.6$ $\mu$m, $l=30$ $\mu$m, $M=130$ kA/m, $\mu_0H=14$ mT, and $\gamma/2\pi=26.5$ GHz/T. 
We believe the small diminishment in the gyromagnetic ratio could be due to other factors and effects not included in the model, such as the boundary interface with the YIG/Py bilayer.
These results confirm the reproducibility of the technique.


\section{Control samples}

\begin{figure}[h]
\includegraphics[width=1.0\textwidth]{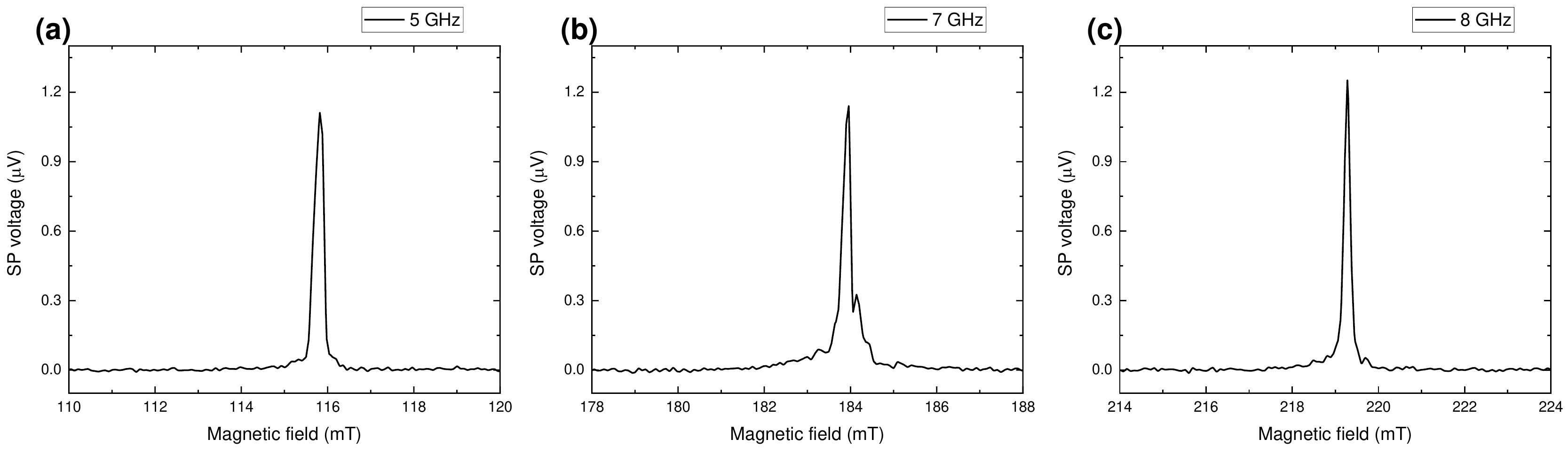}
\caption{\label{fig:amother_remote}
Less intense or not present "secondary" peak in a \textit{Remote} Pt strip on a different 100 nm thick YIG sample at \textbf{a)} 5 GHz, \textbf{b)} 7 GHz, and \textbf{a)} 8 GHz.
Further experiments using rectangular microwave waveguides may avoid surface mode excitations in the bulk YIG film.
}
\end{figure}

\begin{figure}[t]
\includegraphics[width=1.0\textwidth]{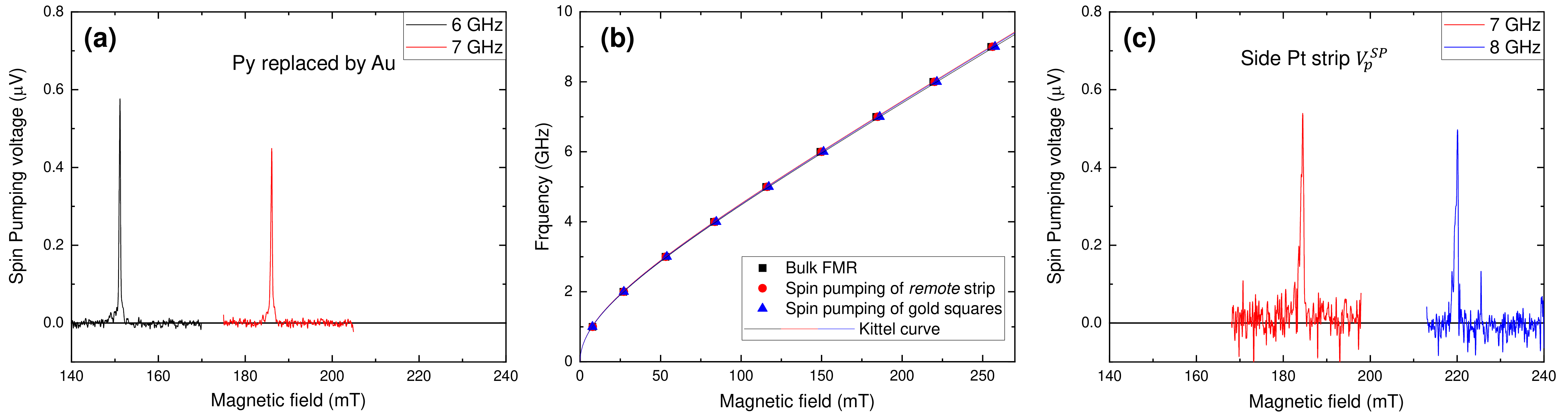}
\caption{\label{fig:Appendix}
\textbf{a)} Absence of the multiple peaks when the Py squares are replaced with gold.
\textbf{b)} Comparison of Kittel fitting between FMR of YIG film, spin pumping voltage in \textit{remote} strip, and when Py square is replaced with Au.
\textbf{c)} Spin pumping voltage with a \textit{proximity} strip, no evidence of cavity resonant modes was observed.}
\end{figure}

In Figure 3 d) on the main text, one can identify a secondary peak at 7 GHz.
This secondary “peak” at the left side of the spin-pumping-FMR peak might be caused by the excitation of the finite \textit{k} bulk-magnons, corresponding to the spin wave modes in the whole YIG sample.
The corresponding peak is also not observed in the FMR absorption.
This secondary structure is less evident or absent in another remote Pt strip using a different YIG sample, shown in Figure \ref{fig:amother_remote}, as well as in Figure \ref{fig:Appendix}.
We believe this structure could be due to the magnon interferences in the bulk-YIG region.
This does not affects the analysis of the resonance modes of the cavity.

Figure \ref{fig:Appendix} a) shows the absence of magnetostatic modes in a control device where an Au film replaces the Py film.
This result rules out the possibility of (multiple) peaks originating from microwave artifacts caused by the proximity of a metallic film to the Pt strip..
Figure \ref{fig:Appendix} b) shows the direct comparison of the Kittel equation for FMR of the YIG bulk, the spin pumping voltage of the \textit{remote} Pt strip, and the spin pumping voltage of the Pt strip when Py is replaced with gold.
 The adjusted Kittel curves show only a slight deviation.
These results also suggest that the magnetic nature of the exchange/dipolar interaction between the YIG and Py films is crucial for constructing the cavity through engineered-design lithography.

The spin pumping voltage was also measured in a Pt strip located near the left side of the Py square, $V_p^{SP}$ in Figure 1 b) of the main text.
However, despite the higher noise level, the spin pumping voltage as a function of the $B$-field did not exhibit any evidence of (multiple) peaks, as shown in Figure \ref{fig:Appendix} c).
It becomes evident that the presence of YIG\textbar Py interfaces on both sides is necessary to create the magnon resonant modes within the cavity.
In Figure 1 c) of the main text, one can identify a Pt strip placed underneath the middle of the Py square.
The electrical resistance of that strip was in the M$\Omega$ range, much higher than the usual $3.5$ $k\Omega$ for 400 nm wide, 35 $\mu$m long, and 8 nm thick Pt strip.
This suggests that the Pt strip may have been damaged during the sputtering and lift-off process involved in the fabrication of the Py square.
As a result, we do not have reliable data for the electrical contact of that particular strip.

\bibliography{aipsamp_sup}